\documentclass[amsmath, amssymb, aps, reprint, groupedaddress, nofootinbib]{revtex4-1}

\usepackage[utf8]{inputenc}
\usepackage{graphicx}
\usepackage{ragged2e}
\usepackage{dcolumn}
\usepackage{bm}
\usepackage{xcolor}
\usepackage{hyperref}
\usepackage[normalem]{ulem}

\newcommand{\mbar}{\overline{m}}

\newcommand{\MSb}{\overline{\mathrm{MS}}}
\newcommand{\df}{{\rm d}}

\def\TeV{\ifmmode {\mathrm{Te\kern -0.1em V}}\else
                   \textrm{Te\kern -0.1em V}\fi}%
\def\GeV{\ifmmode {\mathrm{Ge\kern -0.1em V}}\else
                   \textrm{Ge\kern -0.1em V}\fi}%
\def\MeV{\ifmmode {\mathrm{Me\kern -0.1em V}}\else
                   \textrm{Me\kern -0.1em V}\fi}%
\def\keV{\ifmmode {\mathrm{ke\kern -0.1em V}}\else
                   \textrm{ke\kern -0.1em V}\fi}%
\def\eV{\ifmmode  {\mathrm{e\kern -0.1em V}}\else
                   \textrm{e\kern -0.1em V}\fi}%

\let\gev=\GeV
\let\mev=\MeV

\def\iab{\mbox{ab$^{-1}$}}
\def\ifb{\mbox{fb$^{-1}$}}

\begin{document}

\preprint{CLICdp-Pub-2019-005, IFIC-19-036, IFT-UAM/CSIC-19-122, UWThPh 2019-26}

\onecolumngrid
\vspace*{-1.3cm}
\begin{flushright}
	{CLICdp-2019-005, IFIC-19-036, IFT-UAM/CSIC-19-122, UWThPh 2019-26}
\end{flushright}
\vspace*{-0.3cm}
\title{Top quark mass measurement in radiative events at electron-positron colliders}

\author{M.\,Boronat}
\affiliation{Institut de F\'\i sica Corpuscular (Universitat de Val\`encia/CSIC)}
\author{E.\,Fullana}
\affiliation{Institut de F\'\i sica Corpuscular (Universitat de Val\`encia/CSIC)}
\author{J.\,Fuster}
\affiliation{Institut de F\'\i sica Corpuscular (Universitat de Val\`encia/CSIC)}
\author{P.\,Gomis}
\email[]{corresponding author: pablo.gomis@ific.uv.es}
\affiliation{Institut de F\'\i sica Corpuscular (Universitat de Val\`encia/CSIC)}
\author{A.\,H.\,Hoang}
\affiliation{University of Vienna, Faculty of Physics}
\affiliation{Erwin Schr\"odinger International Institute for Mathematical Physics}
\author{A.\,Widl}
\affiliation{University of Vienna, Faculty of Physics}
\author{V.\,Mateu}
\affiliation{Departamento de F\'isica Fundamental e IUFFyM, Universidad de Salamanca}
\affiliation{Instituto de F\'isica Te\'orica UAM-CSIC\\}
\author{M.\,Vos}
\affiliation{Institut de F\'\i sica Corpuscular (Universitat de Val\`encia/CSIC)}

\date{\today}

	\begin{abstract}
	In this letter, we evaluate potential of linear $e^+e^-$ colliders to measure the top quark mass in radiative events and in a suitable short-distance scheme. We present a calculation of the differential cross section for production of a top quark pair in association with an energetic photon from initial state radiation, as a function of the 
	invariant mass of the $t\bar{t}$ system. This {\it matched} calculation includes the QCD enhancement of the cross section around the $t\bar{t}$ production threshold and remains valid in the continuum well above the threshold. The uncertainty in the top mass determination is evaluated in realistic operating scenarios for the Compact Linear Collider (CLIC) and the International Linear Collider (ILC), including the statistical uncertainty and the theoretical and experimental systematic uncertainties. With this method, the top quark mass can be determined with a precision of $110$\,\mev{} in the initial stage of CLIC, with $1$\,\iab{} at $\sqrt{s} = 380$\,\gev, and with a precision of approximately $150$\,\mev{} at the ILC, with $L = 4$\,\iab{} at $\sqrt{s}= 500$\,\gev. Radiative events allow measurements of the top quark mass at different renormalization scales, and we demonstrate that such a measurement can yield a statistically significant test of the evolution of the MSR mass $m_t^{\rm MSR}(R)$ for scales $R< m_t$.
\end{abstract}

\maketitle

\section[intro]{Introduction}
%
The top quark plays an important role in the Standard Model (SM) and in many of its extensions. The lifetime of the top-quark ($\sim10^{-25}\,\mathrm{s}$) is shorter than the typical timescale to form hadrons. The large decay width cuts off many of the non-perturbative QCD effects in top quark production and decay. Therefore, many top quark properties can be related with high precision to predictions in perturbation theory. That makes the top quark ideal for precise tests of the SM.

The top quark mass and the relevant entries in the CKM matrix are not predicted by the SM and must be determined experimentally. Direct top quark mass measurements at the Tevatron and the LHC experiments have reached approximately $500$\,\mev{} precision~\cite{Aaboud:2018zbu,Khachatryan:2015hba,ATLAS:2014wva} and yield the current direct measurement world average:
\mbox{$m_t = 172.9 \pm 0.4$\,\gev~\cite{Tanabashi:2018oca}}. The experimental uncertainties of direct top mass measurements are expected to improve to approximately $200$\,\mev{} at the HL-LHC~\cite{CMS-PAS-FTR-13-017}. At this time it is not yet clear how the top mass value from direct measurements is related to field-theoretically defined top mass schemes, but theoretical work is ongoing to clarify the relation~\cite{Azzi:2019yne,Hoang:2018zrp}. 
Pole mass extractions from (differential) cross-section measurements at the LHC have quickly improved in recent years. While the latest world average from cross section measurements ($m_t^{\rm pole} = 173.1 \pm 0.9$\,\gev~\cite{Tanabashi:2018oca}) yields a consistent result, recent precision measurements~\cite{Aad:2019mkw,Sirunyan:2019zvx} pose 
some tension with the direct measurements. At electron-positron colliders, a very precise measurement of the top quark mass, with a total uncertainty of around $50$\,\mev{} and a rigorous interpretation of the measurement with respect to mass renormalization schemes, is possible by scanning the center-of-mass energy through the $t\bar{t}$ production threshold~\cite{Abramowicz:2018rjq,Vos:2016til,Gusken:1985nf}.

This letter assesses the potential of future electron-positron colliders to measure the top quark mass in associated production of a top quark pair with an energetic photon. The method we propose reconstructs the differential cross section as a function of the invariant mass $\sqrt{s^\prime}$ of the $t\bar{t}$ system. The radiated hard photon reduces the invariant mass available for top-antitop pair production, and the spectrum develops a strong dependence on the top quark mass, as the 
invariant mass of the $t\bar{t}$ pair approaches the strong interaction production threshold. Such a measurement can be performed in continuum $t\bar{t}\gamma$ production at any center-of-mass energy above the $t\bar{t}$ production threshold and does not require a dedicated $t\bar{t}$ threshold run of the experiment. The extraction of the mass is robust, as it is based on an inclusive $t\bar{t}$ selection, while the 
observable depends only on a reconstruction of the photon energy. The method furthermore offers a rigorous interpretation of the extracted top quark mass in terms of short-distance renormalization schemes, as the differential cross section can be calculated with high precision and the measurement is {\em inclusive} in the top quark decay products.

The analysis is based on a calculation for associated production of a top quark 
pair with an energetic photon radiated from the initial electron-positron pair. We derive a factorization theorem that relates
the $t\bar{t}\,\gamma_{\rm ISR}$  differential cross section to the total hadronic cross section $\sigma(e^+e^-\to t\bar{t}\,X)\equiv \sigma_{t \bar{t}}$, evaluated at the 
invariant mass of the hadronic $ t\bar{t}\,X$ system. The factorization is valid at leading order in the electromagnetic coupling $\alpha_{\rm em}$ and evaluated in the vector-current approximation.
In the region of the invariant mass $\sqrt{s^\prime}$ spectrum where the highest top quark mass sensitivity arises, 
the ISR and the vector-current approximations are valid within a few percent, which is fully sufficient for the purpose of 
this paper. 

For our numerical studies, the description of the vector-current induced total hadronic cross section $\sigma_{t \bar{t}}$ is based on a {\em matched} calculation in analogy to Ref.~\cite{Bach:2017ggt}, that combines an $\mathcal{O}(\alpha_s^3)$ fixed-order (N$^3$LO) description of $t\bar{t}$ production in the continuum~\cite{Hoang:2008qy,Kiyo:2009gb} and a renormalization group improved NNLL calculation for the $t\bar{t}$ threshold region~\cite{Hoang:2013uda}. The latter accounts for bound-state effects and the summation of large logs of the $t\bar{t}$ relative velocity. This approach yields a smooth description of the QCD enhancement of the cross section at the $t\bar{t}$ production threshold due to bound-state effects and the continuum region.

We study the potential for a top quark mass measurement in two concrete scenarios, the CLIC~\cite{Charles:2018vfv,Linssen:2012hp} initial-stage run colliding at a center-of-mass energy of $380$\,\gev{} and the ILC~\cite{Bambade:2019fyw,Behnke:2013lya} run at $500$\,\gev. The statistical uncertainties are estimated for the most up-to-date operating scenarios, which envisage an integrated luminosity of $1$\,\iab{} for the CLIC initial stage~\cite{CLIC:2016zwp} and $4$\,\iab{} for the ILC run~\cite{Barklow:2015tja}. We take into account a realistic experimental acceptance, selection efficiency and resolution, based on detailed simulation studies in Refs.~\cite{Abramowicz:2018rjq,Amjad:2015mma}. We also study the impact of the luminosity spectrum on this measurement in detail.


\section{Observable definition\label{def}}
\begin{figure}[h]
	\centering
	\includegraphics[width=0.95\linewidth]{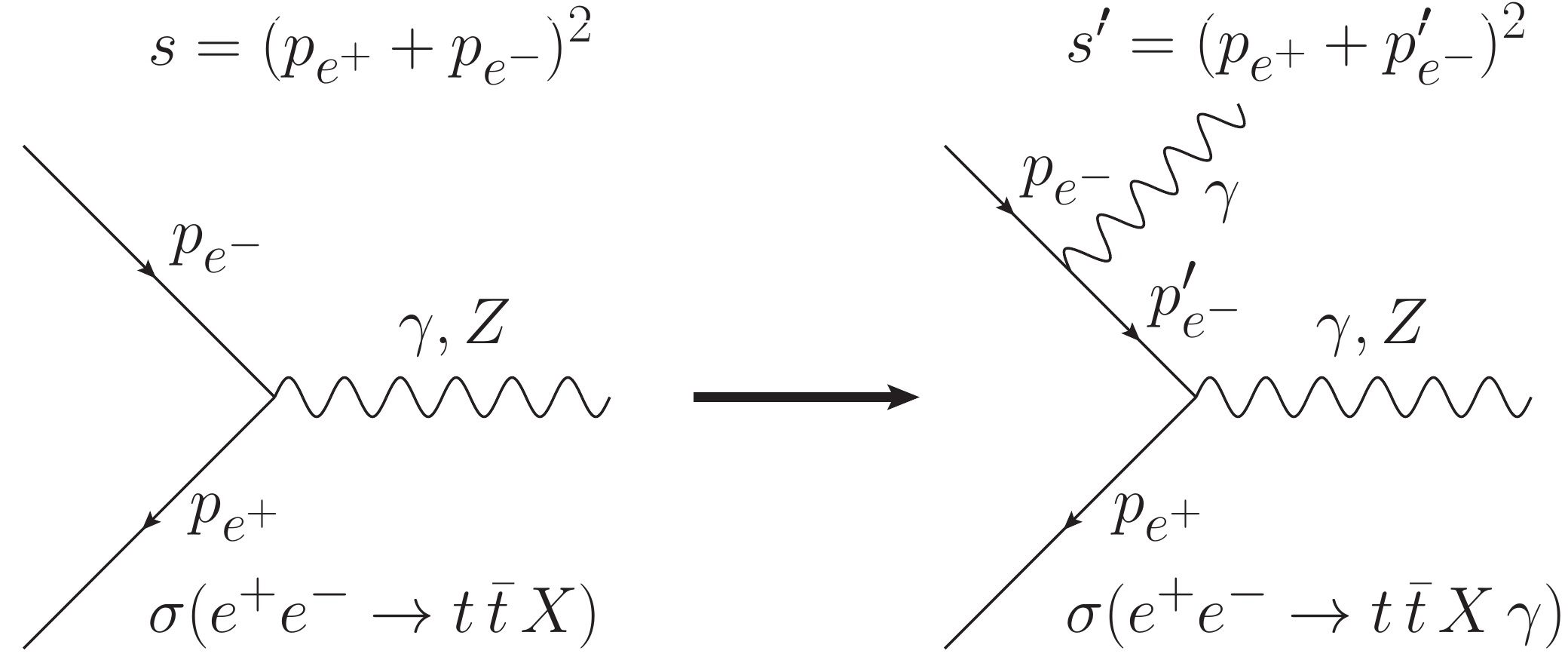}
	\caption{\label{fig:sprimedef} Feynman diagrams representing the center-of-mass energy available for the $t\bar{t}$-pair production before (left) and after (right) the ISR photon emission.}
\end{figure}

We consider radiative $e^+e^- \rightarrow t\bar{t} + X + \gamma$ events, such as those depicted in the rightmost diagram of \autoref{fig:sprimedef}, where a top quark pair is produced in association with an energetic photon. The momentum carried away by the radiated photon reduces the phase space available for $t\bar{t}$-pair production.

\begin{figure}[h]
	\centering
	\includegraphics[width=\linewidth]{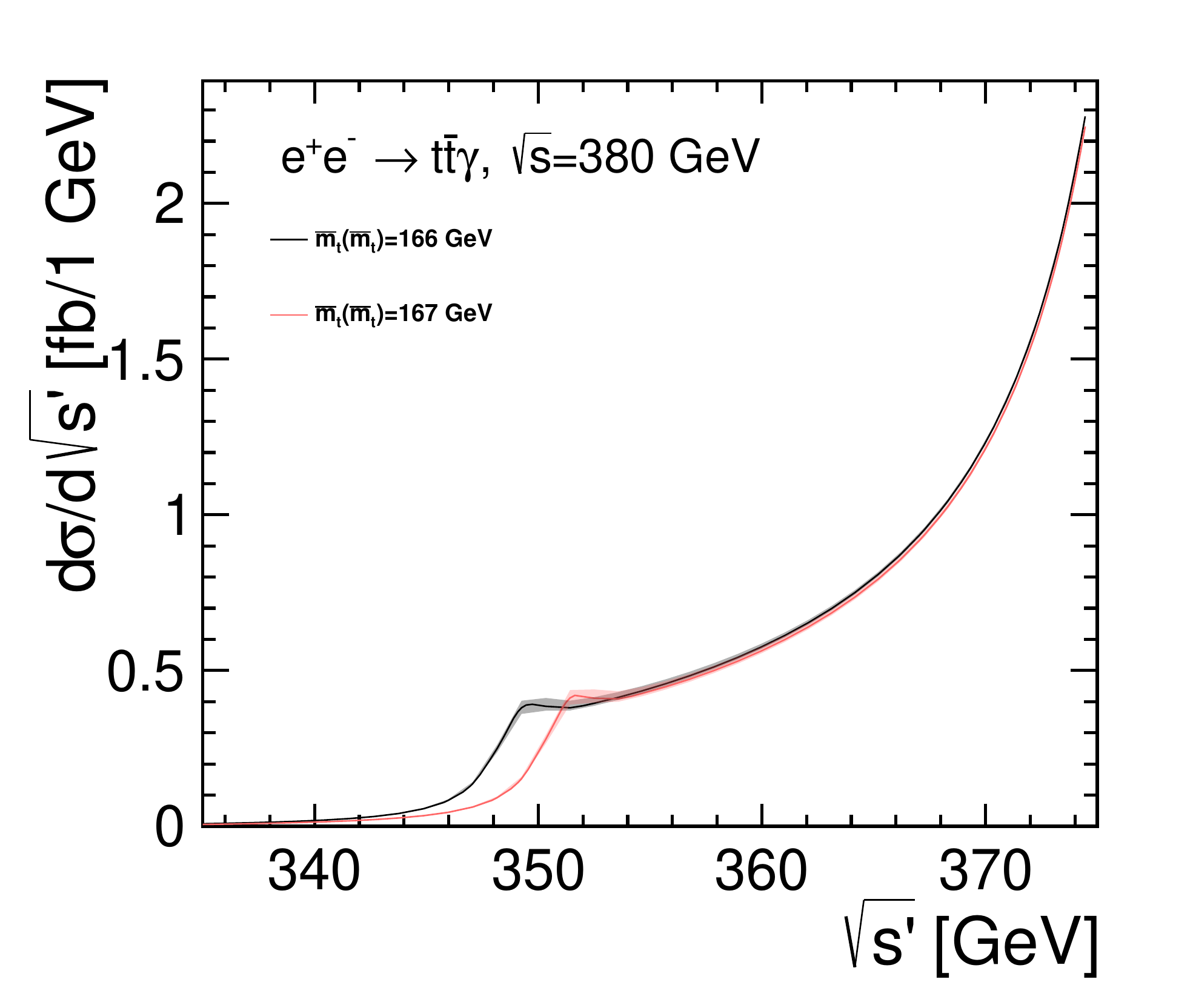}
	\caption{\label{fig:obs} Prediction of the differential cross section versus $\sqrt{s^\prime}$ with the matched NNLL threshold and N$^3$LO continuum calculation for vector-current induced $t\bar t$ production. The black and red curves correspond to two different values of the $\bar{MS}$ mass: $\mbar_t(\mbar_t) = 166\,$GeV and $167$\,\gev, respectively. The photon angle $\theta$ ranges from $8^o$ to $172^o$ and the photon energy $E_\gamma$ must exceed $5$\,\gev.}
\end{figure}

Accounting for the photon emission the squared invariant mass of the $t\bar{t}$ system $s^{\prime}$ is given by:
\begin{equation} \label{eq:sp}
	s^\prime = s \left( 1 - \frac{2 E_\gamma}{\sqrt{s}}\right),
\end{equation}
where $\sqrt{s}$ is the nominal center-of-mass energy of the $e^+e^-$ collider and $E_\gamma$ denotes the photon energy. The differential cross section as a function of $\sqrt{s^\prime}$ is shown in \autoref{fig:obs} for a center-of-mass energy of $380$\,\gev{} and two values of the top quark mass. The maximum sensitivity of the observable is reached at the \mbox{$t\bar{t}$-pair} production threshold (\textit{radiative return} to the $t\bar{t}$-pair production threshold).

Importantly, $s^\prime$ depends only on the nominal center-of-mass energy $\sqrt{s}$ of the $e^+e^-$ collision and the photon energy $E_\gamma$. The center-of-mass energy is precisely known and the photon energy can be measured accurately in the electromagnetic calorimeter system. The determination of $s^\prime$ only depends on these two quantities and is therefore under excellent experimental control. The analysis does in particular not require a detailed kinematic reconstruction of the top quark candidates.

In this method, the top quark mass is extracted from a measurement of the differential $e^+e^- \rightarrow t\bar{t}\,\gamma_{\rm ISR}$ cross section with respect to $s^\prime$. As can be seen in Figure~\ref{fig:obs}, the sensitivity to the top quark mass mainly comes from the region where the photon energy
$E_\gamma$ is close to $E_{\gamma,{\rm max}}=(s-4m_t^2)/(2\sqrt{s})$, the kinematic bound for top quark pair production, where $s^\prime\approx 4m_t^2$. In this region, the dependence of the cross section on $s^\prime$ shows the characteristic 1S bound state resonance enhancement. Physically, this kinematic region corresponds to the photon being radiated back-to-back to a collinear $t\bar t$ pair with $t$ and $\bar t$ having identical momenta. In practice, photons collinear to the incoming electron and positron beams are not accessible experimentally. Therefore, we use the cross section differential in both the photon polar angle $\theta$ and the hadronic system invariant mass
$\df \sigma_{t \bar{t} \gamma}/(\df\cos\theta\,\df \sqrt{s^{\prime}})$, where $\theta$ is defined with respect to the direction of the incoming electron.

%
%
The polar angle $\theta$ is integrated over a range which is limited by the experimental acceptance, typically down to a polar angle $\sim 8^o$. Smaller angles suffer from collinear enhancements that would require resummation. The latter is not included in our theoretical setup but could be accounted for in principle. Soft photon singularities do not arise because we only consider energetic photons with an energy $E_{\gamma} > 5$\,\gev. The highest top mass sensitivity comes from the threshold region, far from this restriction.

\section{Theoretical Prediction}
	\label{sec:theory}

\subsection{Basic approximations and factorization}

As can be seen in Figure~\ref{fig:obs}, an accurate prediction of the $s^{\prime}$ distribution requires a {\em matched} calculation that provides a smooth description valid at the $t\bar{t}$ production threshold and at invariant masses $\sqrt{s^\prime}$ well above threshold. For top mass measurements, the threshold region is essential and requires the best possible precision. For the theoretical prediction we use two essential approximations that significantly simplify the calculations. First, we only include ISR photons, and, second, we only account for vector-current induced top pair production. In the threshold region we have calculated the dominant $t\bar t$ S-wave Wilson coefficient and checked at tree-level that ISR diagrams (which are gauge-invariant) approximate the full Standard Model Wilson coefficient within $2.2\%$ (and $5.2\%$) or better for the cross section integrated over the photon polar angles $\theta$ mentioned above at $\sqrt{s}=380$\,GeV (and $500$\,GeV). In the $t\bar t$ continuum the ISR approximation is known to be still very efficient. We note that the ISR approximation may be further improved by additional photon isolation cuts.

In the ISR approximation and at leading order in the electromagnetic coupling, one can show that the observable factorizes as follows:
%
\begin{align} \label{eq:factorization}
	\!\!\!\frac{\df \sigma_{t \bar{t} \gamma}}{\df\cos\theta\,\df \sqrt{s^{\prime}}} & =
	2\,g(x, \theta)\sqrt{\frac{1-2x}{s}}\frac{\alpha_{\rm em}}{\pi}\,
	\sigma_{t \bar{t}}(s^\prime)+\mathcal{O}(\alpha_{\rm em}^2)\,,\nonumber\\
	x&= \frac{E_\gamma}{\sqrt{s}}\,,
\end{align}
where $g(x,\theta)$ is a calculable function of the photon polar angle $\theta$ and the fraction $x$ of the center-of-mass energy carried by the photon. The derivation of this result is presented in the Appendix 
and the analytic expression for $g$ is shown in Eq.~\eqref{eq:radiative-kernel}.

As an additional approximation we only account for the vector-current induced contribution to $\sigma_{t \bar{t}}$. In the  $t\bar{t}$ threshold region, axial-vector 
induced top pair production is related to P-wave $t\bar{t}$ states, which are suppressed by the square of the top velocity $v$ (defined in the $t\bar t$ center-of-mass frame) with respect to the S-wave contributions induced by the vector-current. The P-wave contributions are known to be at the level of a few percent (see e.g.\ Ref.~\cite{Hoang:2001mm}) 
and therefore negligible at the level of precision of our analysis. In the continuum region, for $E_\gamma>5$\,GeV, the axial-vector induced contribution is still at the few percent level for $\sqrt{s}=380$\,GeV because $v<0.4$. For $\sqrt{s}=500$\,GeV the axial-vector induced contribution in the continuum for $E_\gamma>5$\,GeV can reach around $10\%$. Since the top mass sensitivity of our method arises from the threshold region, our vector-current approximation is viable for $\sqrt{s}=500$\,GeV for the purpose of our analysis. 
For the analysis with real data at $\sqrt{s}=500$\,GeV, however, the axial-vector-current induced cross section contributions must be included.

\subsection{Matched cross section}
To obtain the matched vector-current induced contribution to $\sigma_{t \bar{t}}$ we proceed as follows.
In the threshold region, the calculation must account for the QCD enhancement of the cross section due to
non-relativistic bound-state effects. This entails that the top velocity $v$  is an additional small expansion parameter and terms proportional to $(\alpha_s/v)^n$ [\,and possibly $(\alpha_s \log v)^m$\,] must be summed to all orders in $\alpha_s$. The description of the threshold region used in this work is based on the renormalization-group-improved vNRQCD calculations of Refs.~\cite{Hoang:2013uda,Hoang:2001mm} which include $(\alpha_s \log v)^m$ resummation and achieved NNLL (next-to-next-to-leading logarithmic) precision. The resulting QCD uncertainties are around $5\%$~\cite{Hoang:2013uda} and determine the theoretical uncertainties of our main numerical top mass analysis discussed in \autoref{sec:results}. The precision may be improved in future work through a combination with the N$^3$LO fixed-order results of Ref.~\cite{Beneke:2015kwa}, but at this level also the ISR and vector-current approximations should be lifted by additional dedicated computations.

For the continuum region of the vector-current contribution to  $\sigma_{t \bar{t}}$, where the fixed-order QCD expansion applies, 
we use the exact results from~\cite{Kallen:1955fb} at $\mathcal{O}(\alpha_s^{0,1})$ 
and the Pad\'e approximation results at $\mathcal{O}(\alpha_s^{2,3})$ from Ref.~\cite{Hoang:2008qy}, which are numerically equivalent to the corresponding results of
Refs.~\cite{Kiyo:2009gb,Maier:2017ypu}. The Pad\'e approximation uncertainties are subleading compared to renormalization scale uncertainties for our analysis, where top mass sensitivity arises mainly for $\sqrt{s^\prime}\lesssim 380$\,GeV.

The NNLL threshold results and the fixed-order $\mathcal{O}(\alpha_s^{3})$ cross sections are matched using the approach of Ref.~\cite{Bach:2017ggt}, originally applied to combine the NLL \mbox{vNRQCD} threshold and the $\mathcal{O}(\alpha_s)$ fixed-order cross sections. The matching procedure is based on the sum of both cross sections and minus an `expanded cross section' that takes care of double counted terms. The \mbox{vNRQCD} and expanded cross sections are multiplied with a switch-off function that suppresses them away from threshold, which is crucial to achieve a smooth transition to the fixed-order cross section far away from threshold. The variation of the switch-off function parameters provides an estimate of the uncertainties inherent to the matching procedure. We have extended this method to NNLL and  $\mathcal{O}(\alpha_s^{3})$ and found that the uncertainty of the matching procedure systematically decreases when going to higher orders. At  NNLL and  $\mathcal{O}(\alpha_s^{3})$ the uncertainty of the matching procedure on the cross section is smaller than its renormalization scale counterpart 
in the intermediate regime between threshold and continuum, where the matching procedure matters~\cite{Widl2017,Widl:2018}.

\subsection{Short-distance mass schemes}
\label{sec:theory_mass}

The matched cross section systematically keeps track of the dependence on the top quark mass and the proper renormalization scheme. The 1S~\cite{Hoang:1998ng,Hoang:1998hm,Hoang:1999ye} scheme $m_t^{\rm 1S}$ is employed for the threshold cross section and the MSR~\cite{Hoang:2008yj,Hoang:2017suc} scheme $m_t^{\rm MSR}(R)$ is used for the
fixed-order and the `expanded' cross sections. The latter is matched to the $\MSb$ scheme\footnote{$\MSb$ stands for the modified minimal subtraction scheme. The pole mass is obtained when renormalising the quark self-energy with the on-shell condition, such that fluctuations from all positive energies are absorbed into the mass. In contrast, in the $\MSb$ scheme one defines counterterms such that only the $1/\epsilon$ divergences cancel, which effectively absorbs into the mass definition fluctuations larger than $\mbar_t(\mbar_t)$. The MSR mass interpolates between these two extremes, absorbing fluctuations above the scale R. The name reflects its close relation to the $\MSb$ scheme and R-evolution.} for renormalization scales above the top mass. As the renormalization scale $R$ for the MSR mass we adopt $R=m_t^{\rm 1S}\,v_\star$. The quantity $v_\star$, given in \autoref{eq:softscale}, is a proxy for the top velocity in the $t\bar t$ center-of-mass frame. So the scale $R$ is close to the center-of-mass 3-momentum of the top quarks (which smoothly depends on $\sqrt{s^\prime}$) and the top mass. This dynamical MSR mass scheme choice is very close to the 1S mass in the threshold region. It minimizes the size of the Coulomb singular corrections in the fixed-order results when leaving the threshold region and smoothly connects to the $\MSb$ scheme in the high-energy regime, which minimizes fixed-order QCD corrections there.
The relations of the three mass schemes and the MSR mass R-evolution equations are implemented at $\mathcal{O}(\alpha_s^{4})$ precision using the results of 
Ref.~\cite{Hoang:2017suc} and the $\MSb$ mass $ \overline{m}_t\equiv \overline{m}_t(\overline{m}_t)$ as the matching point between the MSR and the $\MSb$ mass. Throughout this paper we adopt the convention that mass values and uncertainties will be presented in the $\MSb$ scheme $\overline{m}_t$. (More details are given in Refs.~\cite{Widl2017,Widl:2018}.) Note that renormalon issues and the pole mass ambiguity~\cite{Beneke:2016cbu,Hoang:2017btd} do not play any role in our analysis which is exclusively based on short-distance mass schemes. The perturbative uncertainties in the relation between the three mass schemes we employ is at the level of $10$ to $20$\,MeV, which also represents the principal theoretical limitation of a top mass determination based on our method, as far as the conceptual control of the mass schemes is concerned.

An interesting aspect is that measurements of the top quark mass within different $s^\prime$ (or $E_\gamma$) regions are sensitive to QCD dynamical effects at different length scales and thus allow for $m_t^{\rm MSR}(R)$ top mass measurements at different $R$ scales, in analogy to $\mu$-dependent determinations of the strong coupling $\alpha_s(\mu)$ from processes dominated by QCD dynamics at different energy scales~\cite{Tanabashi:2018oca}. While the precise scheme (and renormalization scale) one picks to visualize the resulting `running' parameter measurements has some degree of arbitrariness, an experimental confirmation of the scale-evolution predicted by theory -- once a suitable scheme choice is made -- represents a non-trivial consistency check of QCD. In this context it is also interesting to note that, as was first pointed out in Ref.~\cite{Voloshin:1992wg}, for renormalization scales below the quark mass, the scale evolution of a short-distance quark mass is linear in contrast to scales above the quark mass where it is logarithmic. This is realized also by the R-evolution of the MSR mass $m_t^{\rm MSR}(R)$ matched to the $\MSb$ mass~\cite{Hoang:2008yj,Hoang:2017suc}.

\subsection{Renormalization scale uncertainties
\label{sec:theory_uncertainty}}

The theoretical uncertainties (related to the unknown higher order corrections) are obtained from scale variations of the matched cross section following the approach of Ref.~\cite{Bach:2017ggt}, where the scale multipliers $h$ and $f$ are used to coherently parametrize the variations of the hard ($\mu_H=h\, m_t$),\, soft ($\mu_S=h f m_t\, v_\star$) and ultra-soft ($\mu_{\rm US}=h f^2 m_t v_\star^2$) scales in the threshold cross section (see also Refs.~\cite{Hoang:2013uda,Hoang:2012us}) and the renormalization scale of the fixed-order and expanded cross sections ($\mu_F=h m_t $) around their default scales, where\,\footnote{For the top quark width we use $\Gamma_t=1.5\,$GeV.}:
\begin{equation}
  v_\star =0.05+\Big|\sqrt{\big(\sqrt{s^\prime}-2m_t+i\,\Gamma_t\big)/m_t}\,\Big|\,.
  \label{eq:softscale}
\end{equation}
The multiplier $h$ parametrizes variations relative to $m_t$, and $f$ parametrizes variations relative to $v_\star$. To estimate the uncertainty due to the truncation of the perturbative series, the $h$ and $f$ parameters are varied in the intervals quoted in \autoref{tab:table1}. In this procedure, outlined in Refs.~\cite{Hoang:2013uda,Hoang:2012us}, the combinations of $h$ and $f$ are chosen such that the ultra-soft renormalization scale in the threshold cross section remains within a factor two of its default value, while $h$ is varied between $1/2$ and $2$. This results in a relative uncertainty of $\pm 5\%$ for the cross section in the threshold region~\cite{Hoang:2013uda} and is the basis 
of the theoretical uncertainty in the top quark mass determination of our method.
At this point we stress that our choice of the $R$ scale in the running MSR mass $m_t^{\rm MSR}(R)$ corresponds to a freedom in the choice of a renormalization scheme for the top mass parameter, but is not associated to the theoretical uncertainties, which are fully accounted for by the $h$ and $f$ variations.

\renewcommand{\arraystretch}{1.6}
\begin{table}[h!]
  \centering
\caption{\label{tab:table1}
Variations of the parameters $h$ and $f$ that determine the hard, soft, ultra-soft and fixed-order renormalization scales of the matched calculation. The nominal calculation
corresponds to $h=f=1$ and the variations follow the procedure of Refs.~\cite{Hoang:2013uda,Hoang:2012us}. The resulting $\sqrt{s^\prime}$
distribution is fitted with the nominal calculation and the resulting shifts in the extracted value of $\mbar_t$ are given in the third and fourth rows.  }
\begin{tabular}{c|ccccccccc}
  \hline
  \multicolumn{10}{c}{parameter variations} \\ \hline
$h$ & $1/2$ & $1/2$ & $1/2$ & $1$ & $1$            & $1$             & $2$ & $2$    & $2$\\
$f$ & $2$    & $3/2$ & $1$    & $1$ & $\sqrt{2}$ & $\sqrt{1/2}$ & $1$ & $3/4$ & $1/2$\\
  \hline
   \multicolumn{10}{c}{resulting mass shift $\Delta\mbar_t$\,[\mev] } \\ \hline
$\sqrt{s}= 380$\,\gev: & $-44$ & $-46$ & $-44$ & $0$ & $-1$  & $+8$  & $+29$ & $+30$ & $+45$\\
$\sqrt{s}= 500$\,\gev: & $-55$ & $-58$ & $-54$ & $0$ & $-2$  & $+12$ & $+32$ & $+34$ & $+51$  \\
\hline
\end{tabular}
\end{table}
%

To assess the theoretical uncertainty in an extraction of the top quark mass,
the differential cross section predictions in terms of $\sqrt{s^\prime}$ (integrated over the photon angle) with varied scale settings are fitted with the nominal calculation (for $f=h=1$), where the top quark mass in the latter is floated as a free parameter. The difference of the best-fit mass with respect to the nominal mass is listed in the third and fourth rows in \autoref{tab:table1}, for center-of-mass energies of $380$\,\gev{} and $500$\,\gev, respectively. The perturbative uncertainty in the mass extraction due to missing higher-order contributions is estimated as the envelope of all top mass variations. We find this theory uncertainty\footnote{This estimate is performed at the parton level, for collisions with the nominal center-of-mass energy. The impact of the scale variations may be significantly larger when evaluated at the detector level. Reference~\cite{Abramowicz:2018rjq} finds an uncertainty of $\pm100$\,\mev, nearly double that at the parton level, for the CLIC luminosity spectrum. We assume in the following that the data can be corrected to the parton level before extracting the mass. This could be achieved by sub-dividing the sample using the event-by-event measurement of the visible energy in combination with a sophisticated unfolding of the data using the precisely measured luminosity spectrum. A rigorous demonstration remains for future work.} corresponds to $\pm46$\,\mev{} at $\sqrt{s}= 380$\,\gev{} and $\pm55$\,\mev{} at $\sqrt{s} = 500$\,\gev \,for the $\MSb$ mass $ \overline{m}_t$.

\section{Experimental study}
In this section we present the strategy to take into account experimental uncertainties. We rely on detailed simulation studies by the CLIC and ILC groups~\cite{Abramowicz:2018rjq,Amjad:2015mma} to estimate the size of the most important effects.

\subsection{Event selection}
The top quark pair production mechanism is the dominant six-fermion process for radiation events with a hard photon with
$E_\gamma<E_{\gamma,{\rm max}}$. 
With its striking signature it is readily isolated, reducing the background due to other SM 
processes to the few percent level~\cite{Abramowicz:2018rjq}. Jets are reconstructed with the Valencia Lepton Collider or VLC algorithm, a sequential recombination algorithm for $e^+e^-$ collisions with robust performance in the presence of background~\cite{Boronat:2016tgd}. Jet clustering is exclusive, with the radius parameter $R$ of the algorithm set to 1.6 and $\beta=\gamma=0.8$. The two $b$-tagged jets are identified with the LCFIplus package~\cite{Suehara:2015ura}, a crucial step in the selection. The selection of the lepton+jets final state~\cite{Abramowicz:2018rjq,Amjad:2015mma} is furthermore based on the presence of an isolated charged lepton (electron or muon)\,\footnote{Events with hadronic $\tau$-decays are not considered at this stage.}. Also the fully hadronic final state can be efficiently selected~\cite{SohailAmjad:2014oua}. No full-simulation studies are available for the di-lepton final state. No reconstruction of the the kinematic properties of the top quarks is required for this analysis.

To take into account the selection efficiency, we assume an overall efficiency of $50$\% for $t\bar{t}\gamma$ events where the photon has an energy greater than 5~\gev{} and a polar angle between 8 and 172 degrees. This is definitely a conservative assumption given the results achieved in full simulation studies of the $t\bar{t}$ final state that demonstrate negligible background levels with similar or better selection efficiency~\cite{Abramowicz:2018rjq,Amjad:2015mma,SohailAmjad:2014oua}. The photon reconstruction efficiency, identification and isolation, discussed in the next section, are included in this overall efficiency and not accounted for separately. 

\subsection{Photon reconstruction}
\label{sec:photon_reco}
Energetic photons leave a characteristic electromagnetic shower in the silicon-tungsten EM calorimeter~\cite{Kawagoe:2019dzh} of the ILD~\cite{Abe:2010aa} and CLICdet~\cite{Arominski:2018uuz} detector concepts envisaged for the ILC and CLIC. They are efficiently identified by the Pandora Particle Flow package~\cite{Marshall:2013bda}.

The expected energy resolution of the electromagnetic calorimeter~\cite{Adloff:2008aa} of the linear collider experiments is:
\begin{equation}
	\sigma / E = 0.166/\sqrt{E} \oplus 1.1\,\%\,.
	\label{eq:resolution}
\end{equation}
In the following the $E_\gamma$ distribution will be binned according to this resolution and the result propagated to the $\sqrt{s^\prime}$ distribution.

The most important limitation of the CLIC and ILC detectors for this analysis is the limited coverage in the forward and backward region. The tracking system and the main granular calorimeters extend out to a polar angle of approximately $8^\circ$. We therefore limit the acceptance of the analysis to the fiducial region $8^\circ < \theta < 172^\circ$ ($|\cos \theta\,|< 0.99$). As the distribution of ISR photons is very forward-peaked, this restriction implies a strong penalty. An extension of the acceptance to include the forward calorimeter systems ($4^\circ < \theta < 176^\circ$) doubles the available statistics. An extension to $2^\circ < \theta < 178^\circ$ would even quadruple it. As the forward calorimeters must deal with important background levels without the aid of the tracking system, this possibility requires a detailed study, that is left for future work.   

Finally, as the calculation we use for the current analysis is based on photons from initial-state radiation, the presence of photons from final-state radiation must be suppressed\footnote{In principle, the contribution to the observable of final state photons can be added to the calculation.}. The studies at the stable-particle level of Ref.~\cite{Boronat:2017phd} establish that a combination of cuts on the photon energy (\mbox{$E_\gamma > 3$\,--\,$10$\,\gev}, depending on the center-of-mass energy) and the isolation angle (minimum angle with respect to the nearest particle, $\Omega_{\gamma,i} > 8^\circ$) is very effective to remove FSR photons, while retaining the sensitivity of the measurement. 

The photon energy measurement is the key to the $\sqrt{s^\prime}$ observable. The response of the electromagnetic calorimeter must be calibrated in-situ to achieve a precise control over the photon energy scale. The electron energy response can be calibrated using $Z \rightarrow e^+e^-$ decays with a statistical uncertainty smaller than $1 \times 10^{-4}$~\cite{Blaising:2019}. A direct handle on the photon response is found in radiative $Z$-boson decays (i.e. $Z \rightarrow \mu^+ \mu^- \gamma$). ATLAS and CMS have applied both methods~\cite{Aaboud:2018ugz,Khachatryan:2015iwa}. Following their results, we assign a conservative uncertainty of $1\times10^{-3}$. 

\subsection{Luminosity spectrum}
\label{sec:lumi_spectrum}

\begin{figure}[h]
	\includegraphics[width=\linewidth]{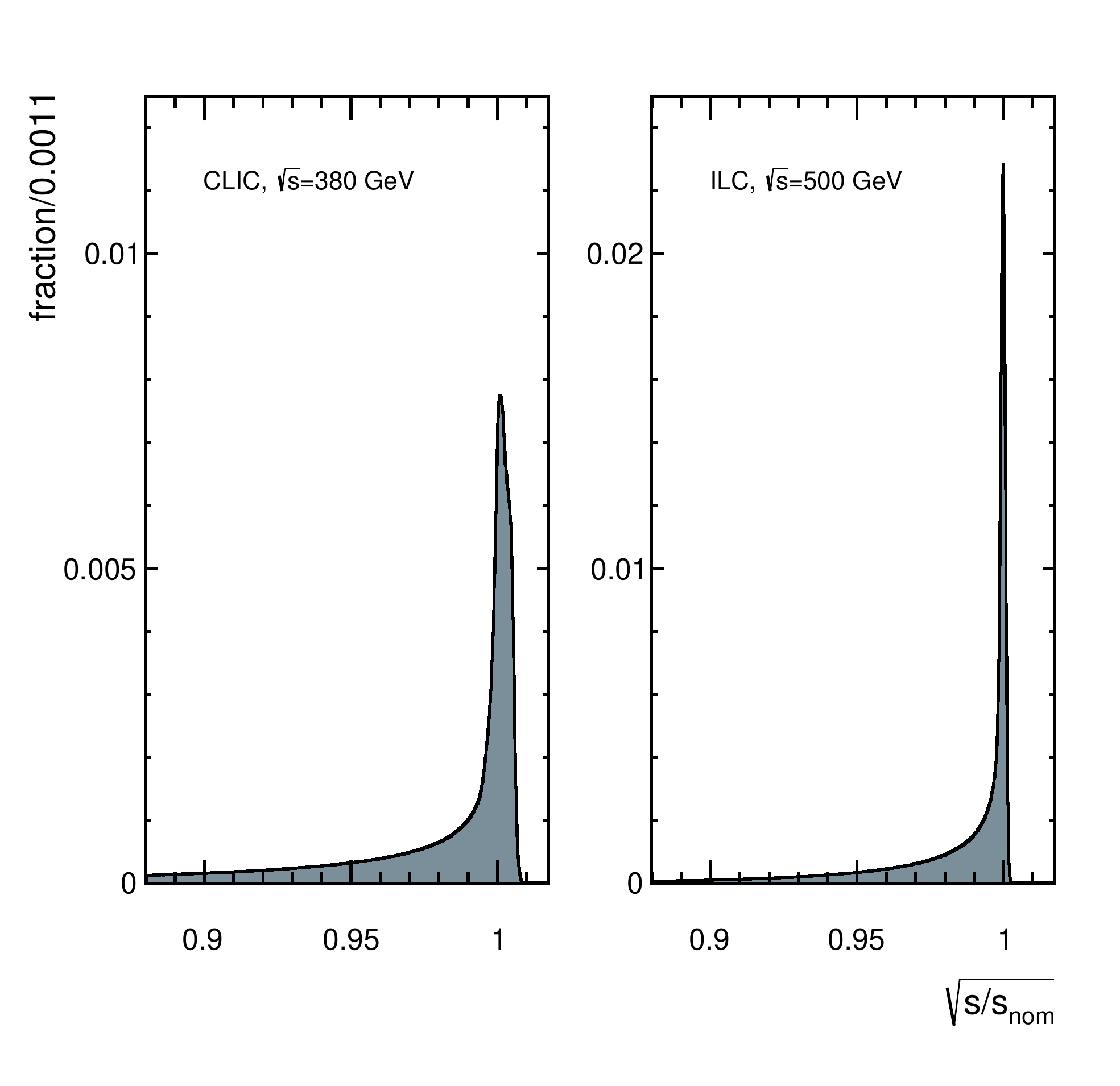}
	\caption{\label{fig:lumispectra} The luminosity spectrum, expressed as the fraction of events per $\sqrt{s/s_{nom}}$ bin, where $\sqrt{s/s_{nom}}$ represents the center-of-mass energy of the collision relative to the nominal center-of-mass energy. The spectra correspond to CLIC at $\sqrt{s}= 380$\,\gev{} (left panel) and the ILC at $\sqrt{s}= 500$\,\gev{} (right panel).} 
\end{figure}

The solid curves in \autoref{fig:obs} assume that the distribution of $e^+e^-$ center-of-mass energy $\sqrt{s}$ is a $\delta$-function at the nominal center-of-mass energy $\sqrt{s_{\rm nom}}$. In practice, beam energy spread leads to a non-negligible width, while beamstrahlung causes a tail towards lower energies called the luminosity spectrum. In \autoref{fig:lumispectra} the expected luminosity spectra for the CLIC run at $\sqrt{s}= 380$\,\gev{} and the ILC run at $500$\,\gev{} are shown, as generated with the GUINEA-PIG program~\cite{Schulte:1999tx} for the nominal accelerator settings.

This luminosity spectrum has an important effect on the observed $\sqrt{s^\prime}$ distribution. Compared to the result obtained with collisions at exactly the nominal center-of-mass energy, the threshold peak gets smeared out substantially. To realize the full potential of this method the effect of the luminosity spectrum must be corrected to good accuracy. In Ref.~\cite{Poss2014} a method was developed to reconstruct the luminosity spectrum in-situ. The Bhabha scattering process $e^+e^-\rightarrow e^+e^-$, with a large cross section, simple final state, and precisely predicted angular distribution, is an ideal calibration process. Following the same approach, we have reconstructed the luminosity spectrum for CLIC380 and ILC500 by fitting a complex functional form to the Bhabha spectrum. A detailed account is found in Ref.~\cite{Esteban:2019}. The uncertainty in the reconstructed spectrum is estimated by varying the parameters of the fit function within their uncertainty. The effect is then propagated to the top quark mass by re-weighting the predicted spectrum with the varied luminosity spectra. 

\section{Results}
\label{sec:results}
\begin{figure*}[ht!]
  \includegraphics[width=0.45\linewidth]{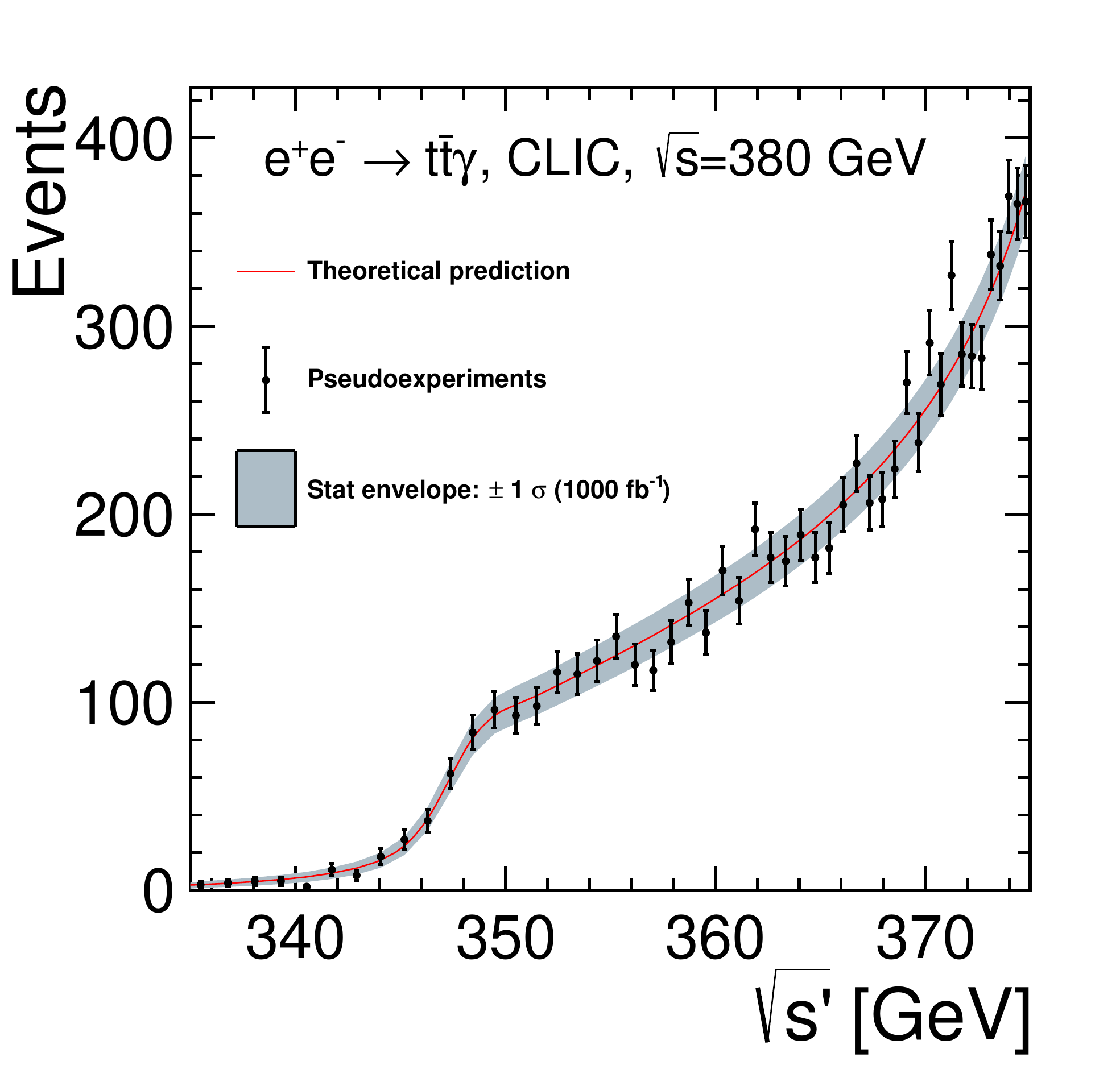}
  \includegraphics[width=0.45\linewidth]{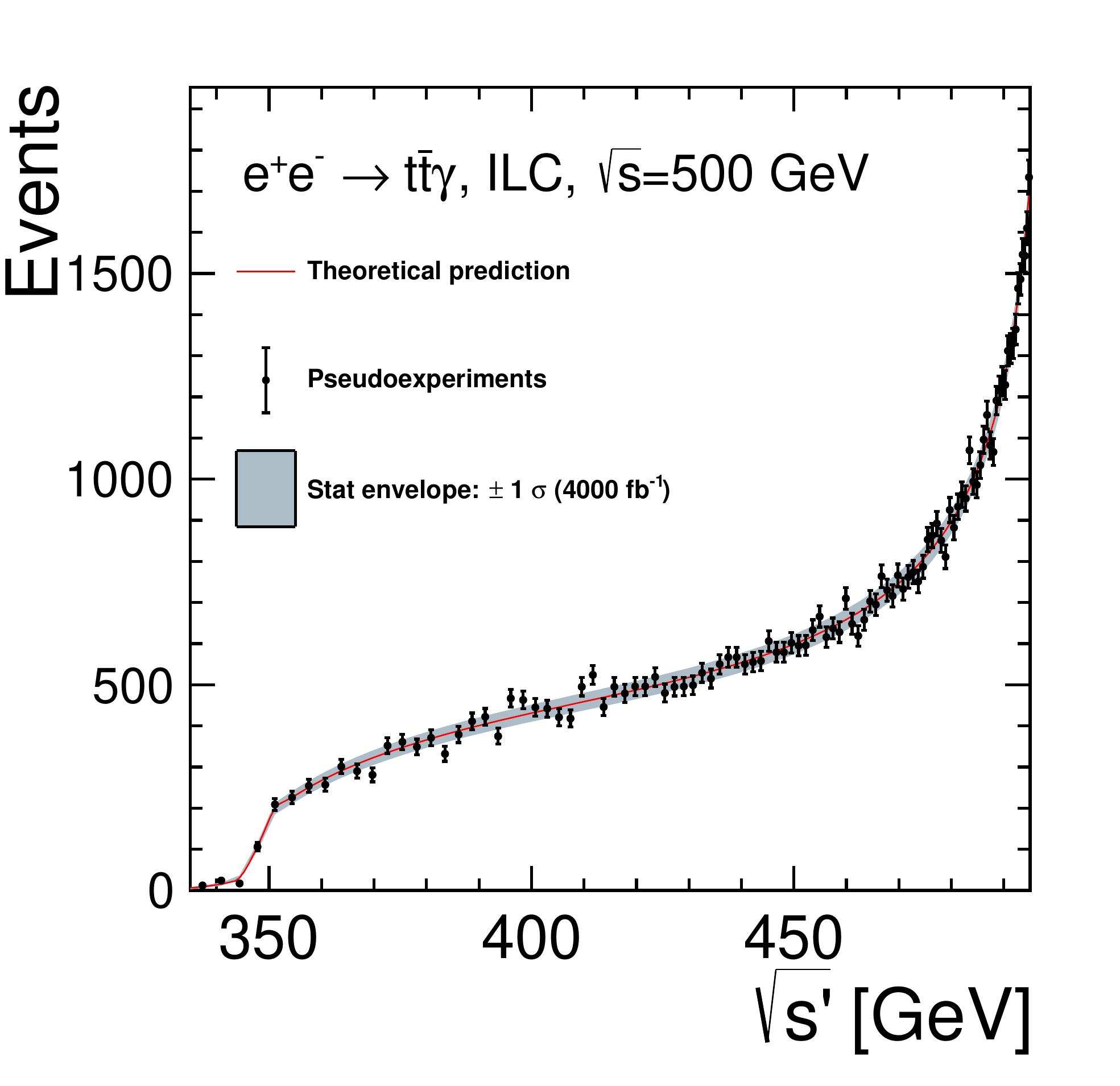}
\caption{\label{fig:pseudoscales} Pseudodata generated with the matched NNLL threshold and N$^3$LO continuum calculation. The left panel shows the result for $\sqrt{s} = 380$\,\gev, with the CLIC luminosity spectrum, the right panel the result for $\sqrt{s} = 500$\,\gev, with the ILC luminosity spectrum. The error on the pseudo-data points and the gray uncertainty band correspond to the $\pm 1\,\sigma$ statistical uncertainty. }
\end{figure*}

The prospects for the top quark mass determination at CLIC and the ILC are estimated using pseudo-data sets. The theoretical cross section predictions 
are collected in $\sqrt{s^\prime}$ bins with a width given by the expected energy resolution of the electromagnetic calorimeter [\,see Eq.~\eqref{eq:resolution}\,]. For each scenario $1000$ pseudo-experiments are generated by applying Poissonian fluctuations around the expected central value. The magnitude of the fluctuations reflects the expected statistical uncertainty, taking into account the production cross-section, integrated luminosity of the official run scenarios and an efficiency of 50\%. The effect of the luminosity spectrum is included by folding calculations at several center-of-mass energies and weighting them according to the luminosity spectrum. Examples of pseudo-experiments  are shown in \autoref{fig:pseudoscales} for CLIC (left) and the ILC (right). The solid curve is the nominal prediction. The error on the pseudo-data points and the gray uncertainty band correspond to the $\pm 1\,\sigma$ statistical uncertainty. 

The pseudo-datasets are fitted to the nominal theoretical prediction with the mass as a free parameter. The statistical uncertainty is estimated as the mean of the uncertainties provided by the TMinuit $\chi^2$ minimization, which is in excellent agreement with the spread of the fitted mass values.

The results are presented in \autoref{tab:uncertainty_breakdown}. We find a statistical uncertainty of $90$\,MeV for the CLIC initial stage at $\sqrt{s}= 380$\,\gev{} and $110$\,MeV for the ILC run at $\sqrt{s}= 500$\,\gev. These uncertainties take into account a realistic estimate of the acceptance, selection efficiency,  photon energy resolution and luminosity spectrum. For the ILC the very large integrated luminosity ($4$\,\iab) is partly making up for the overall smaller statistics in the threshold region. For even higher energies the statistical uncertainty increases even further because of the decreasing statistics in the $t\bar t$ threshold region and the degraded photon energy resolution.

	\begin{table}
		\caption{The expected uncertainty on the top $\MSb$ mass for the runs at $\sqrt{s}= 380$\,\gev{} at CLIC and at $\sqrt{s}= 500$\,\gev{} at the ILC. For both machines, results are presented for an integrated luminosity of $500$\,\ifb{} and for the nominal operating scenario, from Ref.~\cite{Charles:2018vfv} for CLIC and Ref.~\cite{Barklow:2015tja} for the ILC. }
		 \label{tab:uncertainty_breakdown}
		\begin{tabular}{l|cc|cc}
			\hline
			cms energy        & \multicolumn{2}{c|}{CLIC, $\sqrt{s}=380$\,\gev} & \multicolumn{2}{c}{ILC, $\sqrt{s}=500$\,\gev}              \\
			luminosity [\ifb] & $500$      & $1000$     & $500$      & $4000$     \\ \hline
			statistical       &  $140$\,\mev &  $90$\,\mev  &  $350$\,\mev & $110$\,\mev \\
			theory            & \multicolumn{2}{c|}{$46$\,\mev}                          & \multicolumn{2}{c}{$55$\,\mev}                                      \\
			lum.\ spectrum     & \multicolumn{2}{c|}{$20$\,\mev}                          & \multicolumn{2}{c}{$20$\,\mev}                                      \\
			photon response   & \multicolumn{2}{c|}{$16$\,\mev}                    & \multicolumn{2}{c}{~$85$\,\mev}                                      \\ \hline
			total             & $150$\,\mev & $110$\,\mev & $360$\,\mev & $150$\,\mev \\ \hline 
		\end{tabular}
	\end{table}
%
       
The impact of systematic effects on the measurement is estimated by fitting distorted pseudo-data with the nominal calculation and registering the bias in the mass extraction. The theory uncertainty is evaluated by varying the three renormalization scales as described in~\autoref{sec:theory_uncertainty}. The maximum upward and downward variations are taken as the uncertainty. This uncertainty amounts to approximately $50$\,\mev{} in both scenarios and is the dominant systematic uncertainty for the CLIC scenario. The uncertainty of the photon energy response is estimated by varying the energy scale up and down by one per mille. This uncertainty is quite sizable for the ILC scenario, where the photon energy is greater than $100$\,\gev{} for the threshold region. The uncertainty on the luminosity spectrum is propagated as described in \autoref{sec:lumi_spectrum} and is sub-dominant in both scenarios.

\section{Running of the top quark mass}


Measurements of $\MSb$ quark masses at different energy scales have been performed on LEP data for the bottom quark~\cite{Rodrigo:1997gy,Abreu:1997ey} and on HERA data for the charm quark~\cite{Gizhko:2017fiu}. Recently, the CMS collaboration presented the first indications for the running of the $\MSb$ top quark mass at the LHC~\cite{Sirunyan:2019jyn} for scales far above $m_t$. 
As already pointed out in \autoref{sec:theory_mass}, 
the method presented in this letter allows the measurement of a scale-dependent top quark mass for scales below $m_t$  by extracting the mass from different sections of the differential cross section as a function of $\sqrt{s^\prime}$.
As the quark mass scheme to quantify this scale-dependence we adopt the MSR mass $m_t^{\rm MSR}(R)$, where $R=m_t^{\rm 1S}v^\star$ is a proxy for the top quark 3-momenta in the $t\bar t$ center-of-mass frame, which is the relevant dynamical scale for the top mass dependence in inclusive top pair production (see the discussion in \autoref{sec:theory_mass}). 

Here, we present a quantitative analysis to assess the potential of the $500$\,\gev{} run envisaged by the ILC. In principle, the range of energy scales can be further extended by adding higher-energy runs. The $\sqrt{s^\prime}$ distribution is divided in four bins. The first bin with $\sqrt{s'}/\rm GeV$ in the interval $[339,374]$ isolates the threshold region, with its excellent sensitivity to the top quark mass due to the rapidly varying distribution at the 1S resonance peak. For this bin we adopt $R=m_t^{\rm 1S}v^\star(\sqrt{s^\prime}=m_t^{\rm 1S})=25\,$GeV as representative $R$-scale.
Three further bins, $[374,411]$, $[411,445]$ and $[445,495]$, provide mass measurements at higher scales. 
For the second bin $v^\star$ varies by $\pm 20\%$ while for the third and fourth $v^\star$ varies by $\pm 10\%$. The  $v^\star$ variation is sufficiently small so that we can associate the respective mean $v^\star$ value to the representative $R$ value for these bins. This gives $R/\rm GeV=94$, $125$ and $153$ for the second, third and fourth bin, respectively.  
Independent fits are carried out in each bin. The resulting fit values for $\overline{\rm MS}$ mass $\overline{m}_t(\overline{m}_t)$ in each bin are converted to $m_t^{\rm MSR}(R)$ at the respective representative $R$ scale. 

\begin{figure}[ht!]
  \includegraphics[width=\linewidth]{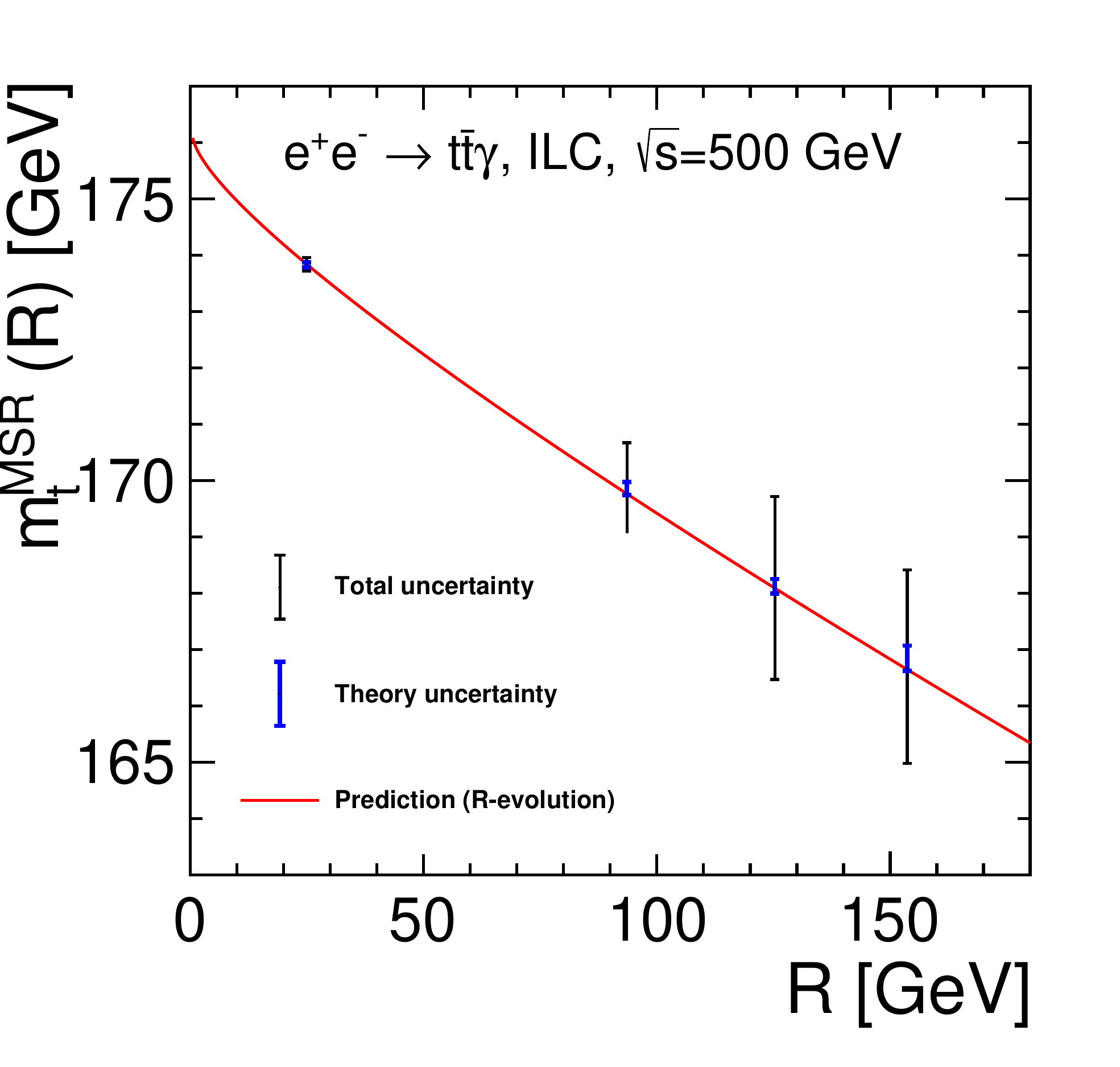}
\caption{\label{fig:running} Prospects for the measurement of the MSR mass evolution with the scale $R$. The error on the pseudo-data points is an estimate of the total uncertainty, which is dominated by the statistical uncertainty.}
\end{figure}

The expected precision of the four MSR mass measurements is indicated in \autoref{fig:running}, assuming the ILC scenario with $4$\,\iab{} at $\sqrt{s}= 500$\,\gev{}. The fit to the bin centered on the threshold region yields a very precise measurement, with a statistical uncertainty of $110$\,\mev{} for $m_t^{\rm MSR}(R=25\,\mbox{GeV})$. At higher $\sqrt{s'}$ the sensitivity degrades very rapidly, and the statistical uncertainty of the mass measurement increases to $0.9$\,--\,$1.7$\,\gev. The theory uncertainty is indicated by blue error bars. It is evaluated at the parton-level by varying the $h$ and $f$ parameters that determine the hard, soft and ultra-soft scales over the range indicated in \autoref{tab:table1}. The theory uncertainty vary from $200$ to $400$\,\mev. Experimental systematic uncertainties are smaller than this.

Taking the precise measurement at the threshold as the reference value, the difference with the results in the other bins at higher $\sqrt{s'}$ ranges from nearly $3$\,\gev{} to over $6$\,\gev{} for the $R$ interval covered by this study. Each of these differences has a significance greater than $3\,\sigma$. The overall significance is greater than $5\,\sigma$. We therefore conclude that this method applied to the $500$\,\gev{} ILC data has the power to demonstrate the R-evolution of the MSR top quark mass $m_t^{\rm MSR}(R)$ for scales $R<m_t$.

\section[theend]{Summary and conclusions}
We present a new method to measure the top-quark mass at electron-positron colliders operated at a center-of-mass energy that exceeds the top quark pair production threshold. In this method, the top quark mass is extracted from the differential distribution of $t\bar{t}\gamma$ events with respect to the invariant mass of the $t\bar{t}$ system, which is related to the photon energy. This method does not require a dedicated threshold scan at $e^+e^-$ center-of-mass energies close to $2m_t$, while it can provide a mass measurement with a rigorous interpretation in terms of a field-theoretical short-distance mass scheme.

The predictions we use for the observable in our analysis are based on a factorization formula, which relies on an ISR approximation, where we account for photon emission from the incoming electron or positron, and on a {\em matched} NNLL+${\cal O}(\alpha_s^3)$ QCD calculation for vector-current induced top production. This calculation provides an adequate description of the QCD enhancement of the cross section in the threshold region where the main top mass sensitivity of our method is located, so that the theoretical uncertainties of the top mass measurement can be estimated realistically, and it smoothly interpolates to the continuum region. For the continuum region our theoretical prediction is less precise, which, however, does not affect the outcome of our error analysis. 

We have assessed the uncertainties that are expected for the CLIC initial stage at $\sqrt{s}= 380$\,\gev{} and the ILC run at $\sqrt{s}= 500$\,\gev. Statistical uncertainties are evaluated for realistic conditions and for the nominal operating scenarios.  We also take into account a realistic estimate of the acceptance, selection efficiency,  photon energy resolution and luminosity spectrum. The theory uncertainty is estimated by varying the renormalization scales in the calculation. A detailed break-down of the uncertainties is presented in \autoref{tab:uncertainty_breakdown}. With relatively modest luminosity, this method can reach a precision of the order of a few $100$\,\mev{} for the measurement of short-distance top quark mass schemes such as $\overline{\rm MS}$ or MSR, improving the top-quark mass measurement beyond the current precision at hadron colliders. For the nominal integrated luminosity ($1$\,\iab) a total uncertainty of $110$\,\mev{} is expected for the initial stage of CLIC at $\sqrt{s} = 380$\,\gev. The ILC run at $\sqrt{s}= 500$\,\gev{} can achieve $150$\,\mev{} with an integrated luminosity of $4$\,\iab.


The precision of this method is not quite competitive with that of a dedicated threshold scan~\cite{Vos:2016til}, but can re-confirm that measurement with partially orthogonal systematic uncertainties. The method we propose can moreover access the top quark mass at various energy scales. Thus, one can test the R-evolution of the MSR top quark mass for scales $R<m_t$ (i.e. the `running' of the top quark mass). The top mass extraction in several $\sqrt{s'}$ intervals with $4$\,\iab{} at $\sqrt{s}= 500$\,\gev{} achieves sufficient precision for the observation with over $5\,\sigma$ significance of the evolution.

Further improvements of the method can be envisaged in several (experimental as well as theoretical) directions. The inclusion of photons detected in the forward calorimeters could extend the polar angle range of the analysis and thus reduce the statistical uncertainty. Even greater gain is possible by including associated production of a top quark pair with a jet initiated by a (final-state-radiation) gluon in the analysis. Progress in the corresponding theory approach to calculate this observable is required to bring it to a level comparable of the present analysis. Among the possible improvements of the theoretical approach we used in the present $t\bar t\gamma$ analysis we count: including both vector and axial-vector induced top pair production, the effects from final-state photon radiation and accounting for electroweak corrections (see e.g.\ Refs.~\cite{Farrell:2006xe,Farrell:2005fk} for corresponding computations for $t\bar t H$ production).

\acknowledgements
%

This result could not have been obtained without the work of the CLIC and ILC collaborations to develop detailed detector concepts, and full-simulation and reconstruction software. MV acknowledges the hospitality of the CERN LCD group. The team at IFIC Valencia is supported by the Spanish national program for particle physics, projects FPA2015-65652-C4-3-R and PGC2018-094856-B-I00 (MCIU/AEI/FEDER, UE), the Severo Ochoa excellence program (SEV-2014-0398) and PROMETEO grant 2018/060 of the Generalitat Valenciana. VM is funded by the Spanish MINECO {\it Ram\'on y  Cajal program} (RYC-2014-16022), the MECD grant FPA2016-78645-P and the IFT {\it Centro de Excelencia Severo Ochoa} Program under Grant SEV-2012-0249, and by the EU STRONG-2020 project under the program H2020-INFRAIA-2018-1, grant agreement no. 824093. AW is supported by the 
the FWF Doctoral Program ``Particles and Interactions'' No. W1252-N2.
AH is supported by the FWF Austrian Science Fund under the Projects No.~P28535-N27 and No.~P32383-N27.

\appendix
\section{Derivation of the Factorization Formula}\label{sec:app}
In this appendix we provide a derivation of the factorization formula shown in Eq.~\eqref{eq:factorization}. We start by relating the total hadronic
cross section to the QCD top quark current $J^\mu$. We assume that quarks are produced directly in the interaction vertex, since secondary
production is not considered in the analysis carried out in this article. We start by noting that the reduced matrix elements $M_C$, where $C=V,A$
denotes the vector and axial-vector current, respectively, according to the leftmost Feynman diagram shown in Fig.~\ref{fig:sprimedef} read
\begin{align}
M_C & = g^C_{\rm ew}(s)\,\overline{v}(p_{e^+})\Gamma^C_\mu u(p_{e^-})\langle X | J_C^\mu(0) | 0 \rangle \,,\nonumber\\
\Gamma^{\{V,A\}}_\mu & =\{ \gamma_\mu\, , \gamma_\mu\,\gamma_5\}\,,\\
J_C^\mu(x) &= \overline{q}_t(x)\Gamma^{C\,\mu} q_t(x)\,,\nonumber
\end{align}
where $g^C_{\rm ew}(s)$ are the electroweak couplings, involving the photon and Z-boson propagators, which are different for vector and
axial-vector interactions (and up-type vs down-type quarks), $u$ and $v$ are the lepton spinors, and $q_t$ is the quantum field for the top quark.
In the second line, the first and second entries in curly brackets correspond to vector and axial-vector currents, respectively. To compute the total
hadronic cross section one has to square the reduced matrix elements (there is no interference between vector and axial-vector currents because we
are totally inclusive), average over the spins of the initial-state leptons, and sum over the polarization and phase space of the inclusive hadronic final state $X$.
Given that we take the initial leptons as massless, the leptonic current is conserved for both vector and axial-vector interactions.
Let us work out the hadronic tensors, defined as the hard matrix element squared, summed over all possible configurations:
\begin{align}\label{eq:hadron-sum}
&W_C^{\mu\nu}(Q)\equiv\nonumber\\
&\sum_X (2\pi)^4\delta^{(4)}(Q^\mu-p_X^\mu) \langle 0 | J_C^\mu(0)^\dagger | X \rangle \langle X | J_C^\nu(0) | 0 \rangle \\
 &=B_C(s) Q^\mu Q^\nu - D_C(s) g^{\mu\nu}\,.\nonumber
\end{align}
Here $Q^\mu$ is the total four-momentum of either the initial or final state, such that $Q^2=s$. The hadronic tensor can be related to the absorptive 
part of the vacuum polarization function. To get the last line we note that $W^{\mu\nu}$ transforms as a rank-two symmetric tensor and use the only
two possible Lorentz structures that can be constructed with the momenta one has at hand. The term proportional to $Q^\mu Q^\nu$ does not
contribute to the total hadronic cross section due to the effective conservation of the leptonic current. For vector interactions, current conservation
of course implies the relation $D(s) = s\,B(s)$. To obtain the total hadronic cross section, we use this result to compute the squared matrix element, 
summed over all final states and averaged over initial polarization (the leptonic trace is the same for both currents), and add the flux factor $1/(2s)$:
\begin{equation}\label{eq:tot-had}
\sigma_{t \bar{t}}(s) = \frac{1}{2}\,\sum_{C=V,A} |g^C_{\rm ew}(s)|^2 D_C(s) \,.
\end{equation}

Next we consider the case of a photon with momentum $p_\gamma$ radiated from the initial state leptons. The momentum flowing through the photon/Z propagator is then $Q^\prime=Q-p_\gamma$, with $s^\prime = (Q^\prime)^2 = s\,(1 - 2x)$ and $x$ defined in Eq.~\eqref{eq:factorization}.
The reduced matrix element, according to the ISR Feynman diagram, as shown on the RHS of Fig.~\ref{fig:sprimedef} reads
\begin{align}
M_C & = g_{\rm em}\, g^C_{\rm ew}(s^\prime)\, F^C_\mu \langle X | J_C^\mu(0) | 0 \rangle \,, \nonumber\\
F^C_\mu & = \overline{v}(p_{e^+})D_\mu u(p_{e^-}) \,,\\
D^C_\mu &=\bigg[\frac{\Gamma^C_\mu (p_{e^-}\!\!\!\!\!\!\!\!\!\slash\;\;\; - p_\gamma\!\!\!\!\!\!\slash\;\;)\gamma_\alpha}{2p_\gamma\cdot p_{e^-}} - 
\frac{\gamma_\alpha (p_{e^+}\!\!\!\!\!\!\!\!\!\slash\;\;\; - p_\gamma\!\!\!\!\!\!\slash\;\;)\Gamma^C_\mu}{2p_\gamma\cdot p_{e^+}}
\bigg] \varepsilon^\alpha(p_\gamma)\,,\nonumber
\end{align}
with $\varepsilon^\alpha(p_\gamma)$ the photon polarization vector. The complete phase space summation now includes also the momenta and helicity $\lambda$ of the ISR photon:
\begin{equation}
\sum_\lambda\! \int\! \dfrac{\df^3 \vec{p}_\gamma}{(2\pi)^3 2 E_\gamma}\sum_X (2\pi)^4\delta^{(4)}(Q^\mu-p_X^\mu-p_\gamma^\mu)\,.
\end{equation}
The sum over $X$ can be performed over the product of the hadronic currents to obtain $W^{\mu\nu}(Q^\prime)$.
Again the $B(s^\prime)$ Lorentz invariant term does not contribute due to the
effective lepton current conservation. The leptonic trace, averaged over the lepton spins and summed over the photon helicities, is the same
for vector and axial-vector currents, and we find
\begin{align}\label{eq:radiative-kernel}
\dfrac{1}{4}\!\!\sum_{s_1,s_2,\lambda}\!\!\! F^C_\mu \,F^{C\,\mu\dagger} &= -\dfrac{4}{x} \, g(x,\cos \theta)\,, \\
g(x,\cos \theta) &= \dfrac{1}{x \sin^2\! \theta}\bigg[1 - 2x + (1+\cos^2 \theta)x^2 \bigg]\,.\nonumber
\end{align}
The radiative kernel $g(x,\cos \theta)$ exhibits both soft ($x\to 0$) and collinear [\,$\theta\to(0,\pi)$\,] singularities. With this result we can
compute the differential cross section:
\begin{align}
\frac{\df \sigma_{t \bar{t} \gamma}}{\df\cos\theta\,\df x} & = \dfrac{\alpha_{\rm em}}{\pi}\, g(x,\cos \theta)\!\!\!
\sum_{C=V,A} |g^C_{\rm ew}(s^\prime)|^2 D_C(s^\prime)\nonumber\\
&= \dfrac{2\,\alpha_{\rm em}}{\pi} \, g(x,\cos \theta) \sigma_{t \bar{t}}(s^\prime)\,,
\end{align}
where in the last line we have used Eq.~\eqref{eq:tot-had} to write the result in terms of $\sigma_{t \bar{t}}$. With a trivial change of variables we obtain the result in
Eq.~\eqref{eq:factorization}. Finally, the radiative kernel can be analytically integrated over an angular bin, and this integrated kernel can be used to obtain the differential cross
section of the events in which the photon ends up in a given cone.

\bibliographystyle{JHEP}
\bibliography{ISR}

\end{document}